\documentclass[a4paper,11pt]{article}
\usepackage[utf8]{inputenc}

\DeclareUnicodeCharacter{0301}{*************************************}

\usepackage{jheppub}
\usepackage{hyperref}

\usepackage{graphicx}
\usepackage{subcaption}
\usepackage[titletoc]{appendix}
\usepackage{xspace}

\usepackage{enumitem}
\setlist{noitemsep}

\usepackage[T1]{fontenc} % if needed

\usepackage[style=numeric-comp,sorting=none]{biblatex}
\begin{document}

% HSF Document Number (only for arXiv)
\noindent
\begin{tabular*}{\linewidth}{lc@{\extracolsep{\fill}}r@{\extracolsep{0pt}}}
 & & HSF-DOC-2022-02 \\
 & & \href{https://doi.org/10.5281/zenodo.7418818}{10.5281/zenodo.7418818} \\
 & & Licence:
\href{https://creativecommons.org/licenses/by/4.0/}{CC-BY-4.0} \\
& & 
\end{tabular*}
\vspace{2.0cm}

\renewcommand{\thefootnote}{\fnsymbol{footnote}}

\title{\boldmath Second Analysis Ecosystem Workshop Report\\
23-25 May 2022, IJCLab Orsay, Paris}

\date{August 2022}

\author{The HEP Software Foundation and IRIS-HEP: \textrm{Mohamed Aly [3], Jackson Burzynski [12], Bryan Cardwell [17], Daniel C. Craik [9], Tal van Daalen [4], Tomas Dado [28], Ayanabha Das [23], Antonio Delgado Peris [10], Caterina Doglioni [3], Peter Elmer [2,b], Engin Eren [27], Martin B. Eriksen [15,16], Jonas Eschle [9], Giulio Eulisse [1,b], Conor Fitzpatrick [3], José Flix Molina [10,16], Alessandra Forti [3], Ben Galewsky [21], Sean Gasiorowski [39], Aman Goel [13], Loukas Gouskos [1,b], Enrico Guiraud [1], Kanhaiya Gupta [41], Stephan Hageboeck [1,b], Allison Reinsvold Hall [32,b], Lukas Heinrich [24,b], Alexander Held [31,b], José M. Hernández [10], Michel Hernández Villanueva [38], Julius Hrivnac [5], Michel Jouvin [5,b], Teng Jian Khoo [33,b], Luke Kreczko [14], Nils Krumnack [22], Thomas Kuhr [7], Baidyanath Kundu [2], Eric Lancon [8], Johannes Lange [18], Paul Laycock [8,b], Kilian Lieret [7], Nicholas J. Manganelli [29], Pere Mato Villa [1,b], Andrzej Novak [20], Antonio Perez-Calero Yzquierdo [10,16], Jim Pivarski [2,b], Mason Proffitt [4], Jonas Rembser [1,b], Eduardo Rodrigues [35,b], Grigori Rybkin [5], Jana Schaarschmidt [4,b], Henry F. Schreiner [2], Markus Schulz [1], Andrea Sciabà [1], Sezen Sekmen [36], Elizabeth Sexton-Kennedy [25,b], Oksana Shadura [34,b], Tibor Simko [1], Nathan Simpson [30,b], Jaydip Singh [26], Nicola Skidmore [3,b], Nicholas Smith [40], Michael Sokoloff [44,b], Graeme A. Stewart [1,a,b], Giles C. Strong [42], Gokhan Unel [37], Vassil Vassilev [2], Mark Waterlaat [43], Gordon Watts [4,b], Efe Yazgan [19]}
 \vspace{5mm}
 \textrm{a\,Lead Editor, b\,Editor}
 \vspace{5mm}
 \textrm{1\,CERN, 2\,Princeton University, 3\,University of Manchester, 4\,University of Washington, 5\,IJCLab, 10\,CIEMAT, 7\,LMU München, 8\,Brookhaven National Laboratory, 9\,University of Zurich, 16\,PIC, 12\,Simon Fraser University, 13\,University of Delhi, 14\,University of Bristol, 15\,IFAE, 17\,University of Virginia, 18\,Universität Hamburg, 19\,National Taiwan University, 20\,RWTH Aachen, 21\,University of Illinois, Urbana-Champaign, 22\,Iowa State University, 23\,Czech Technical University in Prague, 24\,Technische Universität München, 25\,Femilab, 26\,University of Lucknow, 27\,DESY, 28\,TU Dortmund, 29\,University of California, Riverside, 30\,Lund University, 31\,University of Wisconsin–Madison, 32\,US Naval Academy, 33\,Humbold-Universität zu Berlin, 34\,University Nebraska-Lincoln, 35\,University of Liverpool, 36\,Kyungpook National University, 37\,UC Irvine, 38\,Deutsches Elektronen–Synchrotron, 39\,SLAC National Accelerator Laboratory, 40\,FNAL, 41\,Universität Bonn, 42\,INFN, Sezione di Padova, 43\,University of Maastricht, 44\,University of Cincinnati}
}

\abstract{The second workshop on the HEP Analysis Ecosystem took place 23-25 May 2022 at IJCLab in Orsay, to look at progress and continuing challenges in scaling up HEP analysis to meet the needs of HL-LHC and DUNE, as well as the very pressing needs of LHC Run 3 analysis.

The workshop was themed around six particular topics, which were felt to capture key questions, opportunities and challenges. Each topic arranged a plenary session introduction, often with speakers summarising the state-of-the art and the next steps for analysis. This was then followed by parallel sessions, which were much more discussion focused, and where attendees could grapple with the challenges and propose solutions that could be tried. Where there was significant overlap between topics, a joint discussion between them was arranged.

In the weeks following the workshop the session conveners wrote this document, which is a summary of the main discussions, the key points raised and the conclusions and outcomes. The document was circulated amongst the participants for comments before being finalised here.}

\maketitle

\hypertarget{introduction}{%
\section{Introduction}\label{introduction}}

The second workshop on the HEP Analysis Ecosystem~\cite{AEIIWS}
took place 23-25 May 2022 at IJCLab in Orsay, to look at progress and
continuing challenges in scaling up HEP analysis to meet the needs of
HL-LHC and DUNE, as well as the very pressing needs of LHC Run 3
analysis. Since the original
HSF Analysis Ecosystem workshop~\cite{AEIWS}, held in Amsterdam five years ago, 
the ecosystem of
software used for analysis in High Energy Physics (HEP) has evolved
considerably, which was reflected in the preparation and organisation of
the workshop.

To set the scene, looking back to the report~\cite{elmer_peter_2017_6599290} 
from that first
workshop, the central ideas put forward have been vindicated and many of
the inchoate ideas proposed have become a concrete reality. In
particular, the increasing use of Python drove, and was supported by, a
new PyROOT interface~\cite{Galli:2020boj}. Python access to the most powerful machine
learning interfaces from data science is now well supported, with
considerable development of data bridges, e.g., in ROOT~\cite{An:2021} and the widely used
Uproot package~\cite{Pivarski_Uproot_2017}. As these
external pieces have become vital to modern HEP, a suite of packages
that focus on HEP needs to access and use Python data science tools has
grown up, with the Scikit-HEP~\cite{scikit-hep,Rodrigues:2020syo}
project providing a focus for this part of the community. Using ROOT has
been made a lot easier, with a mature Conda installation option and
modularity being planned for new components. Declarative interfaces have
become capable and are being used for analysis at scale, with RDataFrame~\cite{rdataframe}
and Coffea~\cite{Smith:2020pxs,lindsey_gray_2022_7158492} being
the most notable. These interfaces are used in the development of
prototype Analysis Facilities, with backend independence, notebook
interfaces, and many attractive properties in, e.g.,
SWAN~\cite{swan, PIPARO20181071} and Coffea-Casa~\cite{refId0-coffea-casa} and the
development of Distributed RDataFrame~\cite{distrdf} (real metrics for success are
still, however, a work in progress).

The use of continuous integration, encapsulation, code analysis and
workflow management tools has improved in the community, which aids both
scale-out and data preservation.

Although users interact increasingly with Python, as expected, C++
remains wide\-spread and necessary, with no other language really
challenging it in HEP. The performance of I/O in ROOT
\cite{BRUN199781,rene_brun_2019_3895860} has been improved
hugely with RNTuple~\cite{ntuple20}, which beats all other available formats. These
critical low-level improvements link back to Analysis Facility
development and continuing R\&D on different storage interfaces, e.g.,
object stores.

In a few places technology predictions from five years ago did not come
to pass, such as memory resident analyses based on persistent memory
technology or super-high core count CPUs (e.g., Xeon Phi failing to
compete with GPUs). In other areas, such as metadata handling, there was
little general progress although the problem is still seen as relevant
and common between experiments.

Even with these considerable advances in tooling, there remain many open
questions and pressing needs that merited a second workshop. With the
improving pandemic situation, the workshop was organised as a hybrid
event, with 125 people in total registering and more than 70 attending
in-person.

The workshop was themed around six particular topics, which were felt to
capture key questions, opportunities and challenges. Each topic arranged
a plenary session introduction, often with speakers summarising the
state-of-the art and the next steps for analysis. This was then followed
by parallel sessions, which were much more discussion focused, and where
attendees could grapple with the challenges and propose solutions that
could be tried. Where there was significant overlap between topics, a
joint discussion between them was arranged.
\href{https://docs.google.com/document/d/17gPx0eKODpYfJwwler4p-9mR8KllISbDy_LlgOA9tfk/edit?usp=sharing}{{Live
notes}} were taken in all of the sessions. An additional session on
Analysis Workflow Design was held as a breakout. On the final day the
topic conveners gave a brief summary of the key points. This was then
followed by writing sessions, where the outcomes of the workshop started
to crystalise. The final result of that process and deliberation is
given in this document.

The success of this workshop can be attributed to the engagement of
experts from different experiments and projects, and their willingness
to discuss openly. The opportunities afforded by in-person participation
were significant, with many intense and interesting discussions between
participants following on from workshop sessions. The organisers would
like to warmly thank colleagues at IJCLab, particularly Michel Jouvin,
for supporting the workshop at the lab and for arranging a memorable
dinner. Sponsorship for the workshop was provided by IJCLab, CERN-HSF,
IRIS-HEP and Nvidia, to whom we are very grateful.

\hypertarget{topics}{%
\section{Topics}\label{topics}}

\hypertarget{analysis-user-experience-and-declarative-languages}{%
\subsection{Analysis User Experience and Declarative
Languages}\label{analysis-user-experience-and-declarative-languages}}

\begin{itemize}
    \item \emph{Convenors: Jonas Rembser, Alexander Held}
    \item \emph{Speakers: Axel Nauman, Jim Pivarski, Sezen Sekmen}
\end{itemize}

This session focused on the user experience of an analyser performing
the final steps in an analysis pipeline. Analysis typically starts from
centrally produced datasets, which are further reduced by filtering
events or event content. New columns are calculated and added, including
information needed for evaluating systematic uncertainties and
observables required for the analysis. analysers then commonly perform
statistical analysis by building statistical models from template
histograms or directly using data in unbinned models. This workflow was
also discussed in the
``Future
Analysis Workflow Design'' session (see section~\ref{breakout-session---analysis-workflow-design}). The user experience for
statistical analysis was not prominently discussed at the workshop, but
it is connected to the ease of systematic uncertainty bookkeeping. The
second main topic of this track was declarative or domain-specific
languages (DSLs), allowing for the decoupling of physics information
from execution details.

Workshop participants ranked the following three aspects as the
\textbf{top three difficulties} in a typical analysis workflow:

\begin{enumerate}
\def\labelenumi{\arabic{enumi}.}
\item
  \begin{quote}
  \textbf{Systematic uncertainties:} Dealing with systematic
  uncertainties was a recurring topic throughout this workshop and not
  just confined to the dedicated track. There is no uniform way to
  handle systematic uncertainties, and analysers must find solutions
  that work for their specific use case. Difficulties arise from the
  various types of systematic uncertainties (those that alter object
  kinematics are more demanding than weight-based uncertainties) and the
  different approaches to evaluating them. The evaluation may rely on
  additional columns and metadata and may require tools that are either
  centrally provided or implemented by the analyser. See also
  section~\ref{systematics} in this report, which is focused
  on systematics.
  \end{quote}
\item
  \begin{quote}
  \textbf{Metadata:} Challenges related to metadata include finding and
  handling the relevant information, including scale factors or other
  calibration data. Another aspect is bookkeeping: tracking and
  organising all datasets required for the analysis. See also section~\ref{systematics-and-metadata} on metadata.
  \end{quote}
\item
  \begin{quote}
  \textbf{Scale-out:} The move from prototyping to running at scale on
  various sites is referred to as \emph{scale-out}. With varying
  scale-out mechanisms and different environments at each site,
  analysers cannot easily port their analysis from their local machine
  to something running at scale at various facilities.
  \end{quote}
\end{enumerate}

Another aspect deemed crucial to the user experience is
\textbf{interoperability}, particularly for interfaces between analysis
stages. Interoperability between ROOT~\cite{rene_brun_2019_3895860}
and Scikit-HEP~\cite{scikit-hep} tools is
mandatory, as well as the interoperability of key objects like
histograms (e.g., Boost::Histograms in cppyy vs.~pybind11,
boost-histogram vs. ROOT histograms, TH* vs.~ROOT7 RHist). Other
examples are Python bindings, serialised statistical models (e.g., RooFit
workspaces including their JSON version and the
pyhf~\cite{pyhf,pyhf_joss} JSON format for
HistFactory workspaces), and data interoperability at the column-level
(also in memory). Not everything needs to be interoperable, but for
novel ideas to flourish, users should not be locked in one particular
toolset after writing the first code.

A big discussion topic was \textbf{onboarding new analysers} and
first-time user experience. There was broad agreement that documentation
generally has to be improved, but the situation was better than five
years ago at the Analysis Ecosystem
Workshop I~\cite{elmer_peter_2017_6599290}. Several success stories were identified, including the
regularly-updated LHCb Starterkit lessons~\cite{lhcb-starterkit}.
Discussion forums, as well as Slack and Mattermost
channels, are also important for new users. Having a variety of channels
for problem discussion can help engage users with different needs.
Instant messaging platforms allow for quick and informal iterations.
Solutions to problems posted in large channels in instant messaging
platforms can however be difficult to discover for analysers compared to
forum posts or discussions on GitHub. It was also noted that tutorial
writing is challenging: what should be addressed in a tutorial? Feedback
from users can be essential to inform developers what to focus on when
providing documentation material. Dedicated hackathons bringing together
users and developers may be an efficient way of exchanging ideas and
creating new material.

Partially related was the discussion on \textbf{educating analysers in
programming \& software engineering techniques} and \textbf{simplifying
analysis tools}, which often appears contradictory. There was no general
consensus on which direction is more important, as both help to make
analysers more efficient. It was noted that interest in programming
training courses, such as the HSF C++ course, is very high. The
importance of having physicists with advanced knowledge of C++ in
particular was also highlighted, as the personpower for maintaining the
large experiment-specific frameworks is limited. Library developers
noted that users frequently run into relatively basic programming issues
where more formal programming experience would help. A big consideration
regarding the complexity of writing an analysis is how high- or
low-level the interfaces are that analysers are expected to use.
High-level interfaces --- including DSLs --- help users to focus on the
physics and give developers more flexibility for optimising the backends
without changing user interfaces. However, access to lower levels, via
hooks or similar, was seen as crucial to retain the possibility for
users to implement more complex analyses.

The discussion also touched on \textbf{performance}: how can a user
learn that their analysis is not suffering from an easily-avoidable
performance bottleneck? Informal conversations with other analysers were
pointed out as one solution to get a rough feeling for the expected
turnaround time of an analysis. Beyond that, experiments might dedicate
meetings to review techniques and help collect best practices. The
importance of profiling tools was also pointed out. They can not only
help the user with identifying bottlenecks, but their output may also
help experts provide targeted help. In this context, it is desirable for
tools and libraries to expose relevant metrics and performance counters
where possible. Simple tooling to study whether an analysis is
I/O-limited could be helpful. Another critical aspect related to
performance is the analysis implementation time. Time spent writing a
very efficient implementation may surpass all the time saved from the
efficiency.

Declarative or \textbf{domain-specific languages} are an emerging
approach to decoupling physics information from execution details.
General advantages of DSLs include their self-documenting nature and the
decoupling from the backend implementation, allowing updates to the
backend as new technologies become available. One strength of DSLs that
was highlighted is the readability of the analysis description and the
possibility to compare different analyses implemented in the same way.
This can simplify navigating the analysis landscape, simplifying tasks
such as finding overlap in analysis event selections. Several efforts to
develop DSLs for HEP analysis are ongoing, covering both embedded and
external domain specific languages. An embedded DSL is built on top of a
host language like Python, while an external DSL uses its own
interpreter or compiler. Embedded DSLs are more prevalent in HEP,
including FuncADL~\cite{Proffitt:2021wfh} and ROOT's
RDataFrame~\cite{rdataframe} (including abstractions built on top of it). The
Analysis Description
Language Project (ADL)~\cite{adl-project,Sekmen:2020vph} is an example of an external DSL, with
CutLang~\cite{Sekmen:2018ehb} available as an
interpreter. A disadvantage of embedded DSLs is that users might be
tempted to mangle the physics logic with execution details, negating one
main selling point of DSLs. Hence, it is important to continue research
and development on external DSLs in parallel. Furthermore, the existing
embedded DSLs often originate from one specific experiment. Care should
be taken to separate the development of DSLs and tools to abstract away
experiment specific execution details and helper routines to not stall
the development on either side. Finally, as analysis preservation is of
crucial importance to make the most of any physics analysis, it should
be integral to the design of any analysis description and workflow as a
whole. Demanding that users convert their analysis to a reproducible one
after-the-fact is not sustainable.

A \textbf{wishlist for the future} emerged when reviewing discussions at
the workshop, composed of items that we perceive as most important to
address. Related to the handling of systematics, this list includes
automatic optimisation of the analysis' computation graph to only
re-calculate quantities when needed. Some tools, like RDataFrame, are
already designed around this concept, and other tools are encouraged to
focus their R\&D in this direction to ensure sustainable analysis
efficiency. Analysis interfaces should support object facades, meaning
that columns can be grouped into object views for easier reasoning at
the level of physics objects. This can be done by a user-specified
schema or automatic aggregation based on column names. Frameworks such
as Coffea~\cite{lindsey_gray_2022_7158492} and bamboo~\cite{David:2021ohq} are
implementing this already. We recommend the development of standardised
ways to build such object facades to harmonise the access to event
content in different tools. Furthermore, small frameworks or libraries
should help with common chores related to systematics and metadata
handling where possible. The wishlist also includes more documentation
and learning material. Related to this is the demand for more available
Open Data and analyses that use this Open Data to showcase and benchmark
different analysis approaches. Tooling to help analysers with debugging,
identifying performance bottlenecks, and optimising their analysis
pipelines is another point on the list.

\hypertarget{analysis-on-reduced-formats-or-specialist-inputs}{%
\subsection{Analysis on reduced formats or specialist
inputs}\label{analysis-on-reduced-formats-or-specialist-inputs}}

\begin{itemize}
    \item \emph{Convenors: Allie Hall, Jana Schaarschmidt, Loukas Gouskos}
    \item \emph{Speakers: James Catmore, Lindsey Gray, Michel Hernandez Villanueva, Jackson Burzynski, Bryan Cardwell, Lukas Alexander Heinrich, Nick Smith}
\end{itemize}

This session included a review of the design ideas and the status of
reduced formats used in various experiments for previous runs, and
concepts for future runs. It also contained a critical discussion of
cases where specialist, non-standard inputs are needed and how reduced
formats can be adopted to fit these cases. A cross-over discussion was
held to explore questions related to storing and evaluating systematic
uncertainties using reduced formats.

During the MC production chain, or similarly the data processing chain,
several reduction tiers are used, to filter out low level information
that is not needed on the analysis level in a standard workflow.

In ATLAS and CMS, the AOD format is the output from the reconstruction,
holding detailed object information, with a size of about 500 kB/event.
In CMS, this is reduced to MiniAOD~\cite{Petrucciani_2015}, through
slimming and thinning the collections, e.g., removing some tracks and
applying preselection cuts, resulting in a size of 30 kB/event. The next
reduction stage is called NanoAOD~\cite{Peruzzi_2020},
which are flat ROOT ntuples, containing only selected and processed high
level objects stored with reduced precision for floats, which gives 1-2
kB/event. Systematic uncertainties are recomputed on the fly and are
therefore not stored. MiniAOD and NanoAOD serve 85\% of all analyses,
and were in use already during LHC Run-2. The size of NanoAOD is
strictly monitored, and official changes to the format need to undergo
approval. Further information on the CMS computing model is available in
the Phase-2 Computing Model document~\cite{Software:2815292}.

In ATLAS, during run-2, about 100 different reduced formats were used,
tailored for specific physics analyses or combined performance studies.
These formats contained slimmed object collections and entire events
were removed that did not pass some criteria (skimming). This results in
a varying file size typically, but not limited to, 30-50 kB/event, and
event skimming fractions ranging from well below 1\% to above 10\%.
Although these formats were effective for analysis, the huge event-wise
overlap between them when made from simulated events (which tend to have
higher skimming fractions), called for revisions to the model.
Consequently, for Run-3, ATLAS introduces a common format DAOD\_PHYS,
which is not skimmed, and which can serve 80\% of all physics analyses,
with a size of around 30-50 kB/event. A smaller format called
DAOD\_PHYSLITE, primarily aimed at Run-4, has a size of 10-15 kB/event.
The main physics objects in PHYSLITE are calibrated as the format is
made, removing the need to store inputs for calibrations. Currently,
inputs for evaluating systematic uncertainties still need to be stored.
These and other concepts for ATLAS computing are outlined in the HL-LHC CDR~\cite{Calafiura:2729668}.

The reduction chain in Belle II~\cite{belle2computing} is also two-staged. Starting from the RAW format (70 kB/event),
which is the detector output, then comes reconstruction, which is first
reduced to the mDST (mini Data Summary Table) format with 15 kB/event
containing a subset of objects, which is then reduced further to uDST
(user DST) with 20kB/event, which is a skimmed version of mDST but also
augmented with analysis objects. There are about 80 skims in use,
tailored for specific needs. The production of these skims is a
bottleneck in the processing chain in Belle II due to the high I/O load.
Not all analyses can use these skims, but it is important that such
analyses are supported as well, also in light of reproducibility and
long-term data preservation.

Limitations of these concepts were discussed as well. In CMS, about half
of the NanoAODs are customised, which means skimmed or the content
extended (or both), to fit specific analysis needs. There is significant
overlap between these customised formats. A possible solution would be
friend trees, including only the additional information needed for each
analysis. Another possible solution could be the so-called LegoAOD, that
uses central services like ServiceX~\cite{galewsky2020servicex}, Crab~\cite{SPIGA2008267}, 
Dask~\cite{dask} etc., to allow users to
easily add extra columns without having to copy the rest. Another
innovative approach is the use of object stores, which can avoid the
need to copy columns across processing tiers (for example some metadata
is duplicated at every stage of the reduction chain).

It was noticed that the PHYSLITE format from ATLAS is about 5-10 times
larger than CMS NanoAOD. PHYSLITE is currently at a prototype stage,
while NanoAOD are already used in physics analysis. A critical review of
the content of PHYSLITE will lead to further reduction. However, one of
the main culprits is the storage of inputs for systematic uncertainties.
Future R\&D work will therefore explore possibilities to reduce this
information, using external look-up tables, parameterisations or ML
models. Not all analyses require the same level of precision, calling
for a flexible approach for evaluating systematic uncertainties.

While the large majority of analysis can use reduced formats, it is of
crucial importance to look at the remaining cases that need special
inputs. In particular searches for BSM physics and exotic signatures
require non-standard objects, such as for example displaced muons,
disappearing tracks or unique shower shapes, or they rely on low-level
information such as energy stored in each calorimeter layer or even
individual cells. ATLAS also maintains a large list of residual formats
for combined performance work that cannot use PHYS, for example
calibration studies that need jet constituents.

One technical solution was presented, that is to add friend trees that
store additional variables, but only for a subset of events. A case
study for the ATLAS search for displaced jets in the calorimeter was
performed. Simply adding the required topocluster collection to PHYS
would increase its size by about 140\%, while adding a friend tree
holding this collection only for events that pass the trigger increases
the size by just 2\%, which is a very encouraging result.

The focus for future work on reduced formats must be on special inputs
for non-standard workflows, since these will ultimately drive the
storage needs of the experiments.

\hypertarget{machine-learning-tools-differentiable-workflows}{%
\subsection{Machine Learning Tools \& Differentiable
Workflows}\label{machine-learning-tools-differentiable-workflows}}

\begin{itemize}
    \item \emph{Convenors: Nathan Simpson, Lukas Heinrich}
    \item \emph{Speakers: Sean Gasiorowski, Vassil Vasilev, Giles Strong, Engin Eren}
\end{itemize}

Since the first Analysis Ecosystem Workshop in 2017~\cite{AEIWS}, the role of Machine
Learning (ML) in High Energy Physics (HEP) analysis has grown
significantly. Back then, our in-house solution TMVA~\cite{TMVA} was the tool under
discussion at the time, but now usage has shifted almost entirely to
industry-developed tools and automatic differentiation frameworks, which
span methods ranging from boosted decision trees (BDTs), XGBoost~\cite{XGBoost}, and
various neural network architectures.

In terms of applications, the majority of ML approaches are concerned
with \textbf{designing observables} to be used in a standard HEP
inference scheme, e.g., template-based statistical inference concerning
a physics parameter of interest. There is not yet widespread effort to
share these trained (e.g., via self-supervision) representations of
event data to be adapted for downstream tasks. This is in contrast to
fields like computer vision or natural language processing, which have
many ways to distribute pre-trained models, e.g. through
PyTorch Hub~\cite{PyTorchHub},
TensorFlow Hub~\cite{TFHub},
Hugging Face~\cite{huggingface}, and others. This
point warrants further discussion to understand potential utility.

While supervised learning is the dominant methodology,
\textbf{unsupervised learning methods} such as variational autoencoders
are gaining interest for applications such as anomaly detection. More
recently, we have also seen progress in techniques for
\textbf{likelihood-free inference}, such as likelihood-ratio estimation,
which skip the likelihood modelling step entirely in favour of directly
estimating the implicit likelihood of the underlying physics simulation.

From a community development perspective, it is important to ensure an
efficient pathway from new R\&D developments within the HEP ML community
to use within production settings within experiments. To this end, the
availability of \textbf{realistic} and \textbf{openly available datasets
and data simulators} (e.g. fast detector simulation for given
experiments) are important, and ML R\&D work should be encouraged to
assess performance in such more realistic environments.

A difficult tension arises in the context of \textbf{reduced formats}:
ML analysis pipelines may aim to incorporate increasingly lower-level
features (e.g., track or calorimeter data) in order to rely less on prior
fixed representations provided by reconstruction algorithms and optimise
task-specific performance. However, the pressure arising from storage
constraints encourages only the most high-level variables to be
retained. Here, the ideas discussed in the workshop on augmenting
reduced formats with analysis-specific columns may become important.

It was noted that to a large extent the development of ML components
within a physics analysis is \emph{regarded as a separate activity} from
the main analysis development, with its own data preprocessing pipeline,
ML training and evaluation frameworks. As most leading ML frameworks
(JAX~\cite{jax2018github}, TensorFlow~\cite{tensorflow2015-whitepaper}, PyTorch~\cite{NEURIPS2019_9015}, Scikit-Learn~\cite{scikit-learn}) focus on a Python user
interface, the ongoing efforts within the HEP community for
\textbf{python-based analysis workflows} are deemed essential to
overcome this pattern. Additionally, ROOT is maintaining integrations of
major external frameworks into TMVA and on facilities for efficient data
loading from RDataFrame. For integrating finalised ML code into larger
workflows, inference tools, such as ONNXRuntime, lwtnn or SOFIE are
important. As the developments on future \textbf{Analysis Facilities}
take shape, it is important that ML-focused workflows become
well-supported on such resources (experiment tracking, hyperparameter
optimisation, metrics analysis) and some integrations with external
services (e.g. KubeFlow, MLFlow) may be worth investigating.
Additionally, the methods developed for continuous tracking and
evaluation of ML training runs could also be transferred to a generic
continuous evaluation of HEP analyses, and is worth investigating to
test its utility.

During the workshop, there was considerable interest in the emerging
concept of
``differentiable
programming''~\cite{Baydin2020DifferentiablePI} (DP) as a generalisation of the gradient-aware
computation in the ML context. In this scheme, arbitrary numerical
programs can produce not only outputs but also efficiently and precisely
compute gradient information with respect to either program inputs or
program parameters. Gradient information can \textbf{improve ML
applications} by combining malleable ML components with more
domain-specific computation to induce stronger \textbf{inductive bias}
into the model. Similarly, loss functions can be improved by providing
differentiable physics-driven evaluations of the model output as e.g. in
INFERNO~\cite{de_Castro_2019} or
neos~\cite{NEOS}. Moreover,
gradient information of the target function can be used within as
powerful labels as demonstrated in e.g. score-based training approaches
like madminer~\cite{https://doi.org/10.48550/arxiv.1907.10621} and
madjax~\cite{madjax}. Interest in
DP also stems from use cases beyond ML, where e.g. gradient information
may be important for sensitivity analysis with respect to systematic
variations, or improve statistical inference as used for example in
fitting frameworks pyhf~\cite{pyhf,pyhf_joss} and zfit~\cite{Eschle:2019jmu}.
There is also
movement in the direction of end-to-end detector optimisation, including
a recent whitepaper on
the topic~\cite{https://doi.org/10.48550/arxiv.2203.13818}. Communities that have formed from this effort (with a
degree of cross-pollination) include
gradHEP~\cite{gradhep} and the
MODE
collaboration~\cite{MODE:2021yid}.

The current practice of analyses based on a fixed reconstruction scheme
is comparable to standard \textbf{transfer-learning} approaches in ML,
where fixed pre-trained layers are used as inputs to training a
task-specific ``tail'' (i.e.. each analysis tuning selections on fixed
outputs of reconstruction, which was optimised on its own set of
objectives). The pre-trained layers may then be fine-tuned towards the
specific task, using small learning rates, by exploiting the
differentiable nature of the model to re-optimize lower-level features
for a given task (i.e. analyses being able to adjust e.g. low-level
reconstruction parameters). This practice has recently become
commonplace in industry. Developing an \textbf{end-to-end differentiable
HEP pipeline} could enable a similar workflow, however the achievable
improvement is unknown and should be studied in the coming years.

A major hurdle towards DP is the fact that much of the existing code to
analyse HEP data is either not written in differentiable programming
languages or the computations may fundamentally not be differentiable
(e.g. hard categorical decisions, array sorting, ..). For the former,
ongoing developments such as
CLAD~\cite{Vassilev_Clad} or
Enzyme~\cite{10.1145/3458817.3476165} aim to enable AD for
general purpose language and ROOT has reported the intention to leverage
these to integrate differentiation more deeply into e.g. RooFit. On the
latter point, effort has been put into consolidating a set of
differentiable surrogates for common operations in HEP (e.g. histograms)
into a package called
\emph{relaxed}~\cite{relaxed}. These
``relaxed'' operations are smooth analogues to the equivalent hard
operation (e.g. max vs softmax) with a tunable level of approximation.
The relaxed operations can be used during optimisation to find settings
for the pipeline with no approximations, and can also be used for the
final task if desired (possibly incurring a bias, if e.g. using a soft
histogram and assuming Poisson statistics per-bin).

We assessed that more effort is needed to study the real-world benefits
of DP in different applications over other optimisation methods (e.g.
Bayesian optimisation), as well as ensuring that the community at-large
converges on common tooling and methodological efforts for the future.

\hypertarget{systematics-and-metadata}{%
\subsection{Systematics and Metadata}\label{systematics-and-metadata}}

\begin{itemize}
    \item \emph{Conveners: Paul Laycock, Teng Jian Khoo}
    \item \emph{Speakers: Thomas Kuhr, Stephan Hageboeck}
\end{itemize}

\hypertarget{metadata-and-the-data-analysis-wg-metadata-paper-review}{%
\subsubsection{\texorpdfstring{Metadata and the Data Analysis WG
Metadata Paper Review
}{Metadata and the Data Analysis WG Metadata Paper Review }}\label{metadata-and-the-data-analysis-wg-metadata-paper-review}}

In the report from
the first Analysis Ecosystem workshop~\cite{elmer_peter_2017_6599290} in 2017, the first item under
``Missing pieces'' is:

\begin{itemize}
\item
  \begin{quote}
  Easy access to bookkeeping information and other metadata. Common
  support for this across experiment boundaries.
  \end{quote}
\end{itemize}

It is fair to say that nothing happened in the intervening 5 years.

The main outcome from the Data Analysis Working Group Metadata paper
review~\cite{Khoo:2022pja}, prepared for
this workshop, was that the authors had identified a major problem for
HEP: Metadata; and the panel recommended the community follow up on this
important topic. This provoked a discussion at the workshop -- who would
follow up? Judging from the progress made in the last 5 years, the
problem is not about to be resolved. Metadata is a large topic touching
many areas, so would benefit from having HSF take responsibility for it
and systematically follow up on the diverse challenges it presents. An
\textbf{HSF Metadata workshop} is proposed, where stakeholders from the
experiments will be invited to participate and discuss the best way to
organise followup within the HSF community. The importance of leveraging
the existing connections to other HSF working groups and activities was
strongly emphasised during the AE2 workshop.

Using the HSF Metadata workshop as a catalyst, a comprehensive list of
use cases for metadata will be collected in an \textbf{HSF Metadata
Report}, delivered shortly after the conclusion of the workshop. The
report will cover the end-to-end lifecycle of metadata and the data they
pertain to, addressing issues like metadata scopes, evolution and change
of ownership during this lifecycle. Often metadata producers are not
responsible for metadata usage in analysis and these orphaned metadata
can become problematic. The HSF Metadata Report should cover all use
cases, not restricted to analysis, to be able to identify such gaps and
ensure metadata systems cover every metadata source for its entire
lifecycle. The report will detail the use cases and a derived set of
requirements, together with a more complete description of metadata
scopes. The paper could also try to look at commonality across use cases
and where common HEP solutions could be applicable given sufficient
buy-in from the experiments.

The definitions of metadata scopes should address the questions posed by
the panel:

\begin{itemize}
\item
  \begin{quote}
  Metadata is data about data. Which data?
  \end{quote}
\item
  \begin{quote}
  On what does the metadata depend?
  \end{quote}
\item
  \begin{quote}
  Is the metadata known at the time of data production?
  \end{quote}
\item
  \begin{quote}
  How is the metadata produced and by whom?
  \end{quote}
\item
  \begin{quote}
  How is the metadata used and by whom?
  \end{quote}
\item
  \begin{quote}
  Who is the owner of the metadata, and are there transfers of
  ownership?
  \end{quote}
\item
  \begin{quote}
  What relations between different levels of metadata exist and how are
  they handled?
  \end{quote}
\end{itemize}

The review panel gave more feedback useful for the HSF Metadata Report.
While they greatly valued the work of the Data Analysis Working Group
Metadata paper~\cite{Khoo:2022pja} authors and found the paper provided a good
overview of several existing approaches, the guidelines lacked
sufficient technical details to fully represent ``technical
specifications'' for future analysis metadata system implementations. A
certain technical elucidation of guidelines would be desirable, and will
also greatly facilitate the overall understanding of metadata use in
view of analysis preservation and reuse. This detailed understanding
could then help future metadata system designers find commonality while
keeping flexibility to extend metadata schemata to cover particular
specific scenarios. A concrete example of progress in this area may be
the CERN Analysis Preservation framework that uses composable JSON
schema to describe analysis steps that allow both common and particular
JSON fields. A detailed report from the panel will be made available in
the coming months.

\subsubsection{Systematics}\label{systematics}

While recognised as an essential component of experimental analysis, the
book-keeping and processing of systematic uncertainties was considered
the largest pain point in analysis software. The step from coding the
nominal analysis workflow to including all uncertainties is large: a
na\"{i}ve loop repeating all work for many variations inflates CPU and disk
footprints; while avoiding this demands a complex optimisation process
requiring tracking of all and only affected columns of data.

ATLAS is developing algorithm sequences for object corrections that
includes optimisation logic for generating collections with variations
applied and filtered for quality, but variation tracking and optimised
handling is otherwise a task for analysis frameworks, including
RDataFrame and Coffea. There is convergence in ATLAS and CMS on
standardised APIs for extracting systematic uncertainties from metadata
repositories (ATLAS `CP Tools' and CMS CorrectionLib), but these are
considered less applicable in LHCb and other experiments that are
heavily reliant on data-driven background estimates or very
analysis-specific procedures for determining uncertainties. The ATLAS
model has the benefit of an integrated representation of uncertainties,
allowing classification during event processing, and preset groupings of
relevant/recommended systematics on a per-object-collection basis. CMS
has entirely decoupled the application of systematic corrections from
the nominal object corrections, which offers a much more nimble workflow
with minimal software dependencies and a low CPU load. This decoupling
requires that uncertainties be parameterised rather than propagated via
alterations of the nominal procedure, so has implications on
experimental workflows beyond the pure technical solution.

The data dependencies for object systematics encourage the use of an
object-based APIs, but the increasing popularity of columnar analysis
methods suggests that these be satisfied through object facades, which
can be constructed e.g. in RDataFrame. Frameworks can then cleanly apply
variations and track affected columns through downstream operations.

Recommendation:

In conjunction with other ongoing innovations, a streamlined analysis
workflow incorporating systematic uncertainties can be proposed, arising
from an overall consensus at the workshop: Analysis should begin from
fully calibrated, reduced data formats (NanoAOD, DAOD\_PHYSLITE) derived
from the fully reconstructed AOD data. The nominal analysis workflow
should be complemented by a loop handling systematic variations computed
on the fly to favour CPU utilisation over heavier I/O, and with the
framework outputting analysis outputs, which might be histograms or a
small number of final discriminants for all variations.

This model offers space for collaboration on solutions to various common
problems:

\begin{itemize}
\item
  \begin{quote}
  An experiment-agnostic set of labels for systematic variations that
  can be used for tracking during analysis computations, as well as for
  downstream interpretation (visualisation/statistical analysis). This
  could facilitate inter-experiment statistical combinations, and be
  used not only for object/event uncertainties but also to represent MC
  and background uncertainties of relevance to the wider experimental
  community.
  \end{quote}
\item
  \begin{quote}
  A standard API for applying systematic variations to object
  collections, utilising the labels described above. Underlying
  implementations of access to metadata stores and calculations can
  remain experiment-specific, but a common representation also serves
  analysis description and preservation purposes well.
  \end{quote}
\item
  \begin{quote}
  Generalised analysis frameworks or other schemes handling optimisation
  logic, such as the RDF::Vary approach, propagating varied columns
  through all downstream analysis operations.
  \end{quote}
\end{itemize}

As a motivation for and test of these common libraries, an analysis
challenge is proposed, comprising a simultaneous top quark cross-section
analysis between ATLAS, CMS and ideally LHCb. This could share MC event
generation, to undergo simulation/recon\-struc\-tion in the respective
experimental frameworks, and subsequently work from the standard
analysis formats, but share code where possible. To encourage active
participation from the collaborations, the main deliverables would be
the establishment of a more coherent organisation and tracking of
uncertainties in CMS and progress with decoupling uncertainties from
nominal corrections in ATLAS. It is proposed that this challenge be
followed up and reported on in the HSF Data Analysis Working Group.

\hypertarget{real-time-onlinetrigger-level-analysis}{%
\subsection{``Real-time'' online/trigger-level
analysis}\label{real-time-onlinetrigger-level-analysis}}

\begin{itemize}
    \item \emph{Conveners: Giulio Eulisse, Mike Sokoloff}
    \item \emph{Speakers: Caterina Doglioni, Daniel Charles Craik}
\end{itemize}

``Real-time'' online and trigger-level analysis (RTTA) can encompass
anything from monitoring detector performance to reconstructing physics
objects for persistence for later data analysis to producing histograms
(perhaps multidimensional) that can be used directly in published
analyses (see \cite{RTAworkshop2019} for an introduction to the wide scope). A key challenge for the immediate future is ensuring that
trigger software and offline software produce the same
(indistinguishable) results, even if executing on significantly
different hardware. For example, LHCb's Run 3 first level software
trigger (Allen, \cite{Aaij2020}) executes on Nvidia GPUs in real time. What level of
difference is tolerable when running on different hardware, and what
tests do we need to demonstrate that we can tolerate the differences?
This is critical for generating and analysing simulated data in WLCG
facilities. In the case of LHCb, Allen provides an x86 back-end that
produces what are judged to be tolerably indistinguishable results.
While Allen compiled for x86 is around 4x slower than equivalent
hand-optimised x86 reconstruction code, this overhead is irrelevant in
the overall LHCb context. Experiments require some re-calibration of
trigger-level data prior to, or during, analysis workflow. Simulating
trigger-level hardware using WLCG resources requires the same level of
validation. An interesting question is how WLCG will provide
heterogeneous resources that reproduce those used in triggering in the
future, which is probably an inevitable trend, albeit not one primarily
motivated by this use case. These could include Nvidia GPUs (used by
LHCb), AMD GPUs (used by ALICE), and a variety of FPGAs (used at some
level by all experiments).

In the longer term, ``real-time'' triggers might be able to use elastic
resources in addition to the bespoke resources dedicated to individual
experiments. LHC experiments typically record data during 5 x
10\textsuperscript{6} seconds per year, and they do not acquire data
every year. To the extent resources sit idle (or mostly idle) for
extended periods of time, they are wasted. A possible alternative would
be moving (some) data from the experimental halls to elastic resources
for higher level triggering. This requires that the bandwidth be
available for transporting the data and that the elastic resources
(potentially CPUs, GPUs, and FPGAs) be available when needed. Such
solutions were investigated in the context of Run 3 but judged to be
unaffordable due to the cost of network infrastructure required. Careful
studies of technical, financial, and administrative (policy) issues will
therefore need to be done before such a model can be seriously
considered for the future. The potential benefits include more efficient
use of hardware (and personnel) and the possibility to dramatically
expand the computing power available to the highest level triggers on
short notice. A serious downside is the outsourcing of quality assurance
work together with associated sociological and management overheads,
given that it has to date generally proven to be difficult if not
impossible to capture and predict all requirements of HEP data processing
chains in specification documents.

\hypertarget{analysis-facilities}{%
\subsection{Analysis Facilities}\label{analysis-facilities}}

\begin{itemize}
    \item \emph{Convenors: Oksana Shadura, Nicole Skidmore}
    \item \emph{Speakers: Robert Gardner, Alessandra Forti, Lindsey Gray, Nick Smith, Enric Tejedor Saavedra}
\end{itemize}

\hypertarget{building-analysis-facility-infrastructure}{%
\subsubsection{Building Analysis Facility
Infrastructure}\label{building-analysis-facility-infrastructure}}

Analysis facilities (AF) can be broadly described as the infrastructure
and services that provide integrated data, software, and computational
resources to execute one or more elements of an analysis workflow. These
resources are shared among members of a virtual organisation and
supported by one or multiple organisations.

At the workshop, the community identified the following key areas of
analysis facilities that are of interest for further investigation:
\textbf{federated identity management, modern data management, and data
delivery techniques, resource-sharing mechanisms, and efficient methods
for sharing user environments} and many others. In this section a brief
explanation of each of the key areas and ideas discussed at the workshop are
summarised for further consideration by the community.

We expect that this document will help the recently founded the
\textbf{HSF Analysis Facilities Forum}~\cite{hsfaff} working group to provide
a generic list of requirements for analysis facility providers and
architects, providing best practice guidelines that will include the
design and key features provided by future AFs as well as a list of
possible research topics to be investigated later.

\hypertarget{interoperability}{%
\paragraph{Interoperability}\label{interoperability}}

Architects of AFs should provide interoperable solutions, meaning that
users should be able to navigate seamlessly from one Analysis Facility
to another with an extensible and modular design to accommodate future
needs without disruptions.

We would also like to acknowledge the importance of simultaneously
supporting ``leg\-acy'' analysis methods within the same computing
environment (or site, facility) to facilitate adoption from the end-user
physics community. This has the added benefit of conveniently
introducing and benchmarking new methods and providing realistic
feedback.

\hypertarget{identity-management-integration-in-analysis-facilities}{%
\paragraph{Identity management integration in Analysis
Facilities}\label{identity-management-integration-in-analysis-facilities}}

One of the most highly demanded features is the integration of federated
identity management in Analysis Facilities~\cite{broeder2012federated}, which allows using an
account from one facility to create an account and log in to a different
facility and related authorisation schemes including various projects
and account resource allocations.

The integration of Federated Identity management is not only limited to
AFs. The same problem arises when we look at HPCs and external
cloud-based resources. WLCG/OSG have to be made aware that with highly
dynamic and interactive work this problem gets more urgent. Identity
management is the main building block, but in addition, also
authorisation and accounting have to be federated.

Over the workshop, it was also highlighted multiple times that we need
to provide the "best practices" on how to integrate tokens and
federated identity into a new ecosystem. This includes a discussion of
the integration differences between the WLCG.

\hypertarget{data-organisation-management-and-access}{%
\paragraph{Data organisation, Management and
Access}\label{data-organisation-management-and-access}}

In terms of DOMA-related topics, the majority of the community is
concerned about the transparent movement of data between Analysis Facilities
and other storage systems used within HEP.

Fast access to input data is one of the most important aspects of an AF.
The access can be from local storage or can be remote. Users need to be
able to find the data as they would on any other type of experiment
resources and to be able to access these resources in a timely manner.
The output of the analysis needs also to be shareable between AFs and
with other sites. To satisfy these requirements an AF is expected to be
integrated with the experiments' Data Management systems and the
experiments' authentication and authorisation system (see Federated
identity section). The access to should be as fast as possible, so it is
expected that AFs, which have a high degree of repeatedly accessing the
same data, will have a system of local caches, in particular XRootD
based XCaches~\cite{fajardo2020creating} have the advantage of being fully integrated with ROOT,
the dominant file format, thus they are able to transfer only portions
of files, reducing the volume of transfers. The size and the QoS of the
caches should be evaluated.

New workflows may also use different data formats for performance
reasons or because they are more compatible with columnar analysis.
Evaluation of such formats is ongoing in the experiments and
transformation services such as ServiceX~\cite{galewsky2020servicex} are envisaged to give the users the capability to transform data on the fly from one format to another.
ServiceX functionality and scalability needs to be tested.

Another aspect we discussed during the workshop was the type of storage.
Experiments applications and services have been designed to use POSIX
file systems but the emergence of object stores with their scalability
and efficiency in serving data raises the possibility that new workflows
can be adapted to use distributed object stores and experiments would
need to review their requirements for POSIX. Whilst it is not for the
AFs to define this at a general level, AFs themselves might support different types of
storage, so this should be investigated.

%Leaving this out for the Latex document
%(e.g. the way Nick Smith mentioned in his \href{https://indico.cern.ch/event/1125222/contributions/4875668/attachments/2449795/4198186/ncsmith-objectstores.pdf}{{presentation}} on the AE2 workshop)

\hypertarget{resource-sharing}{%
\paragraph{Resource sharing}\label{resource-sharing}}

During the IRIS-HEP
Blueprint meeting~\cite{IRISHEPblueprint} an idea was mentioned to create a multi-site
substrate project which would federate contributions from multiple
resource providers (institutes and public cloud), offering a flexible
platform for service deployment at the scales needed to test the
viability of system designs and closely matching the concept of
"infrastructure as code", in the case of Kubernetes~\cite{k8s} - i.e. the substrate as
the medium upon which infrastructure (services) can be deployed
declaratively and flexibly and also be easily redeployed on the other
facilities. For efficient resource sharing, the community expects the
integration and co-location of analysis facilities with existing centers
and proposes the adoption of Kubernetes as a substrate to maximise
resource contributions and utilisation. This will also require
understanding how computing resources sharing will be organised,
including, importantly, the sharing of storage space among Analysis
Facilities.

Another important feature of scalable resource sharing is maintaining a
clear separation between execution engines and the resource sharing
layers, which means being able to switch between notebook provisioning
(fully interactive resources), analysis facility execution frameworks
(e.g. Dask workers~\cite{dask}), and traditional batch and SSH login environments.

An R\&D topic that should be investigated is how to implement fair
sharing in a scalable environment, especially the provision of resources
suitable to the type of analysis the user wants to perform. A possible
solution that was discussed is to design an intelligent portal,
knowledgeable of the different capabilities at various Analysis
Facilities, rather than users having to discover or keep track of this
information themselves.

\hypertarget{sharing-environment}{%
\paragraph{Sharing environment}\label{sharing-environment}}

The main requirement for users while doing collaborative physics
analyses is the ability to share code and results.

We discussed how environments can be shared efficiently, using existing
mechanisms, such as singularity containers distributed on~\cite{cvmfs}, and how
we could share instead of standardise environments by providing sharing
mechanisms for ``data analysis'', ``ML training''. Instead of trying to
replicate the same environment at different sites and force everyone
to use the same, the best approach could be to implement a sharing
mechanism, i.e. an infrastructure that can support any Data Analysis or
ML environment the users may require. No analysis can be performed with
a single environment as different parts of the workflow often have
different requirements (e.g. analysis-oriented environments such as
LHCb's lb-conda~\cite{lbconda}, powering environments to perform PID calibration
studies). Further interesting mechanisms allowing more efficient sharing
of environments and data are export mechanisms to HEPData~\cite{maguire2017hepdata}, sharing
cached (short-lived) data, and exporting metadata for analysis
preservation.

A key problem is how to preserve user environments when moving from one
facility to another. To solve this problem we discussed existing
mechanisms, such as centrally managed software stacks (e.g. LCG
releases~\cite{villanueva2019building}). Another opportunity could be to provide container images that
contain a tested combination of libraries that can be used to perform
the standard analysis workflows, or to allow users to ``bring-your-own''
image to take advantage of non-standard libraries using BinderHub~\cite{binderhub} for
Jupyter environments~\cite{jupyter} and workers. There needs to be a preservation
strategy for these container images and maintenance of their security
and authorship providence (software metadata).

\hypertarget{best-development-practices-for-analysis-facility-architects-and-developers-using-kubernetes}{%
\subsubsection{Best development practices for Analysis Facility
architects and developers using
kubernetes}\label{best-development-practices-for-analysis-facility-architects-and-developers-using-kubernetes}}

Currently, higher-level analysis systems are being developed within a
Kubernetes environment, and the role of having a flexible (programmable)
the substrate was noted. Since the time of the workshop, Kubernetes (Cloud
Native Computing Foundation~\cite{cncf}) has emerged as a promising technology for this role. Already we have demonstrated deployments of processing frameworks for columnar data
(Coffea~\cite{Smith:2020pxs}) and data delivery (ServiceX~\cite{galewsky2020servicex}), plus identity management combined
into declarative deployments providing an "analysis facility"
(Coffea-Casa~\cite{refId0-coffea-casa}). The identified requirements are flexible access to
infrastructure resources (CPU, disk, GPU and I/O), the declarative
nature of deployments for reproducibility at multiple centers and
providing the capabilities for processing frameworks and interactive
interfaces.

Similarly, we need to work on standardising analysis facility deployment
tools (today Helm~\cite{helm}, tomorrow its successors or the new tools) and
supporting the surrounding infrastructure (Helm chart catalogs,
container registries). This defines a new training area for system
administrators, to build a culture of analysis facility providers
experienced in the dependent technologies. Beyond deployment, this
includes a new set of troubleshooting skills and infrastructure
knowledge (pods, containers, controllers, etc.).

For novel analysis systems and related services, we also need to draw the
clear distinction with site-specific attributes and removal of
dependencies (e.g., a site's identity or account management service,
LDAP, etc.).

A ``best practice'' that the community recommends for future analysis
facilities is that developers should always ask themselves about
portability. For instance, could a facility be deployed at different
institutions? As well as the scalability of deployment, could an
analysis facility be used by more than one user community with the
experiment-specific dependencies removed, or the needed hooks properly
abstracted? Another question is if the deployment pattern lends itself
to open source best practices and whether it can be deployed with
confidence by a K8s cluster operator without significant supervision and
what priors are assumed if this is not the case?

From a security point of view, for analysis facilities built with one or
more internet-facing services, image security is paramount, and services
for vulnerability scans, patch updates, etc. need to be addressed.

We should also encourage analysis facility developers during the design
phase to have in mind the following capabilities: privileged interfaces
for user management, roles, and what metadata is needed. A public-facing
dashboard using open technologies with status reports about the AF such
as performance metrics and community metrics (such as analysis groups,
tasks, users, institutions) will help to measure impact.

Over the course of the workshop, we discussed the differences between
Kubernetes "distributions" (e.g. vanilla Kubernetes vs OKD~\cite{okd}) and how to
get applications to work on both without the users having to choose.
There was an agreement that there should be guidelines for K8s apps
developers that take into account the differences. The K8s application code
adjustments needed, depending on new Kubernetes releases, is another
the challenge that developers will need to keep pace with and could be
efficiently resolved through better interaction with the Cloud Native
community~\cite{cncf}.

\hypertarget{tracking-analysis-performance}{%
\subsubsection{\texorpdfstring{Tracking analysis performance
}{Tracking analysis performance }}\label{tracking-analysis-performance}}

To provide an extensive overview of how resources are used and to ensure
that an informed decision about which AF to use can be made, Analysis
Facilities should publish some key parameters related to performance
(occupancy, etc.). For this reason, we need to provide instrumentation to
understand user modalities, preferences, and analysis throughput. This
can result in a variety of metrics - session/task timings, resource
consumption, and contextual system "business" markers (the single-use
systems, many users/groups, undersubscribed and oversubscribed
resources).

Examples of metrics that could help to have an insight into how and if
efficiently used resources in the analysis facility:

\begin{itemize}
\item
  \begin{quote}
  workflow ID,
  \end{quote}
\item
  \begin{quote}
  CPU, RAM, swap,
  \end{quote}
\item
  \begin{quote}
  I/O (local storage and network),
  \end{quote}
\item
  \begin{quote}
  software stack,
  \end{quote}
\item
  \begin{quote}
  job failure rate,
  \end{quote}
\item
  \begin{quote}
  Time To Completion (TTC),
  \end{quote}
\item
  \begin{quote}
  data source - entirely local or cached from a Data Lake and formats
  used on input (PHYS, PHYSLITE, AOD, MiniAOD, NanoAOD, NTuple,etc..),
  \end{quote}
\item
  \begin{quote}
  formats written (columns), ratios.
  \end{quote}
\end{itemize}

A cost-effective or user-effective solution could require users to test
jobs on subsamples of the data prior to the final submission on the full
samples. This option is, for example, already explored by the LHCb
experiment (exploiting GitLab pipelines) for NTuple productions with the
analysis productions framework. This validation step allows one to
collect many of the metrics mentioned above.

We need to introduce a concept of telemetry for various components of
Analysis Facilities that will provide an overview of what users are
actually doing and which components of the analysis pipeline are run.
Telemetry in the software stack will likely require a dialog with
software analysis framework developers to coordinate a common approach
as well as to establish an API for tools to write their own usage logs
and metrics. Most likely a standard set of Kubernetes tools can do this,
along with some log message format standards.

Collecting more data could also allow the optimisation of resources
requested for known workflows. This means that we could, for example,
scale user jobs for heterogeneous architectures, particularly using
accelerators in a more adaptive way and to try to tune the analysis
configurations in a heuristic way through some validation phases.

The IRIS-HEP Analysis Grand Challenge (AGC)~\cite{agc} workflow defines an analysis
benchmark that could be easily re-implemented and executed on any
generic Analysis Facility and help to showcase how to use existing
analysis facilities on a scale appropriate for this analysis. During the
workshop the IRIS-HEP Analysis Grand Challenge (AGC) workflow was
suggested as one of the analysis facility performance benchmarks, giving
the possibility for AFs architects or resource managers to understand
better data analysis bottlenecks at AF. AGC also could be used to
explain to the users of the analysis facility what it could look like
for an efficient data analysis pipeline.

\hypertarget{collecting-user-requirements-and-metrics-of-success-for-analysis-facilities}{%
\subsubsection{Collecting user requirements and metrics of success for
analysis
facilities}\label{collecting-user-requirements-and-metrics-of-success-for-analysis-facilities}}

Every project, including Analysis Facilities, needs to determine
the success of a project and help project managers to evaluate a project's
status, foresee risks, and assess the quality of work. For this reason it
should be initially envisaged to perform regular user surveys and define
key parameters to capture the success of analysis facilities. As a
baseline to evaluate AF success as a project, we could propose to use
the standard key metrics such as the total number of users or a number of
analyses or the evolution of usage over time, both in terms of the number of
users and compute time spent. Other examples of metrics could include
the number of users per week versus time or the number of active users
within a given time window and by which method users are interacting
(interface, traditional batch), the total number, volume, and format of
files consumed/written and shape of workflow statistics.

Another important category of metrics is the more user-oriented
metrics, such as the number of reported issues, the time to answer
tickets, the time taken to solve the issue, feedback from the user
satisfaction surveys, etc., as well as other user engagement metrics
such as quality and user satisfaction using prepared documentation,
tutorials and quality of user-engagement training for a given facility.

We also need to continue to improve analysis facilities by collecting
additional user requirements and keeping track of improvements over time
and the uptake over time. This also could be defined as an additional
set of metrics.

\subsection{Breakout Session - Analysis Workflow
Design}\label{breakout-session---analysis-workflow-design}

During the workshop, there was a breakout session on \emph{Analysis
Workflow Design}, the summary of which is linked
\href{https://docs.google.com/document/d/116ITweumV9AKIBvMn-GcPQ4UQ--xvZLhAgtjlgGhZdc/edit?usp=sharing}{{here}}.

\hypertarget{conclusions-outcomes}{%
\section{Conclusions / Outcomes}\label{conclusions-outcomes}}

Here we summarise the main outcomes of the workshop that should be
followed up:

\begin{enumerate}
\item
  Object facades, which group columnar data into views that allow
  reasoning about physics objects, are important for a good analyser
  user experience. The HSF Data Analysis Working Group should coordinate
  a harmonisation effort to ensure consistent behaviour across different
  tools in this regard.
\item
  The interoperability of tools should be strongly encouraged and the
  HSF should encourage appropriate discussions between development
  teams.
\item
  Best practice for onboarding new analysts (in analysis models and
  programming techniques) is a crucial area in which to have continued
  commitment between the HSF Training and Data Analysis groups and the
  experiments.
\item
  Continued R\&D for reduced formats should focus on how to accommodate
  special inputs for non-standard workflows. Possible approaches include
  the use of friend trees to augment existing reduced formats or using
  object stores to access columns across processing tiers.
\item
  The curation of open datasets (or equivalently simulators) along with
  benchmarks for common problems in machine learning is of clear
  importance, and should be made a focus of the experiments that have
  sufficient data and meaningful task definitions to suit this format.
  This will bring the field more in-line with industry standards for
  replication, give a better feel for how well new methods perform on
  certain tasks, and reduce friction for non-physics experts to
  contribute in a meaningful way to particle physics research.
\item
  Differentiable programming received a clear interest from those
  present in the workshop, but lacks thorough comparisons with existing
  methods at scale, and requires concentrated effort to harmonise
  software and methodology efforts. This should be a priority going
  forward for those involved to make it clear to the field when, how,
  and if these methods warrant usage.
\item
  Metadata requires dedicated follow up and an HSF Metadata Workshop is
  proposed, tentatively for early 2023. This should be coordinated by a
  team nominated by HSF Coordination. The product of that workshop
  should be an HSF Metadata Report, covering all Metadata use cases
  through their end-to-end lifecycle. Discussions on how best to provide
  dedicated follow up, via new or existing working groups or otherwise,
  will also be an important part of the workshop.
\item
  Concerted effort is needed to tackle systematic uncertainty handling,
  identified as the largest pain point in the analysis workflow.
  Agreement on an idealised workflow for analysis of calibrated, reduced
  analysis formats provides a model for R\&D and a set of deliverables.
  A joint top-pair cross-section analysis on open MC between ATLAS/CMS
  and LHCb was proposed as a test challenge for development of common
  tools to describe and extract standard uncertainties both in analysis
  code and in downstream interpretation frameworks.
\item
  The future of computing is moving towards higher concurrency and
  increased use of accelerators. We believe that analysis usage is no
  exception to this trend. Many of the tools discussed at the workshop
  should be supported to continue the work of adapting to this future
  reality, and taking full advantage of facility resources that offer
  accelerators like GPUs and FPGAs. That is clearly an element of the
  path towards our ultimate goal of reducing time to insight, by making
  results available within interactive time scales. The HSF, through its
  various working groups, should attempt to foster best practice and
  knowledge sharing in this transition.
  \begin{enumerate}
  \item
    As a particular specific aspect, the numerical stability and
    reproducibility of multi-architecture codes should be studied, with
    an understanding of when sufficient and acceptable architecture
    independence of results has been achieved.
  \end{enumerate}
\item
  Community involved in development modern AFs identified the following
  key areas of analysis facilities that are of interest for further
  investigation and improvements: Identity management integration,
  modern data organisation solutions and heavy usage of data delivery
  techniques, investigate the object stores, the resource sharing
  mechanisms and the efficient user friendly methods for sharing user
  environments between analysis facilities and many others.
\item
  With more analysis level services deployed within the Kubernetes
  environment an essential area for improvement is standardisation of
  analysis facility deployment tools and support of the surrounding
  infrastructure. Defining a new training area for system administrators
  will give analysis facility providers sufficient experience in the
  dependent technologies.
\item
  To provide an extensive overview on how AF resources are used and to
  ensure that an informed decision about which AF to use can be made, it
  would be useful if Analysis Facilities publish key metrics related to
  performance and usage. It is initially envisaged to perform regular
  user surveys to capture the success of analysis facilities, focusing
  on user-oriented or user engagement metrics. We can also use this to
  understand user modalities, preferences and analysis throughput to
  guide AF development.
\end{enumerate}

\section{Acknowledgments}
We thank the Irène Joliot-Curie Laboratory, the HEP 
Software Foundation, IRIS-HEP, and nVidia for sponsoring
this workshop.
We especially thank the administrative staff of the
Irène Joliot-Curie Laboratory for their help.
This work was supported in part by the U.S. National Science Foundation (NSF) Cooperative Agreement OAC-1836650 (IRIS-HEP) as well as the German Research Foundation (DFG) under grant EXC-2094-390783311.
We acknowledge support of participants from their
national funding agencies, including:
CNRS/IN2P3 (France);
INFN (Italy);
DFG (Germany);
STFC (United Kingdom);
DOE and NSF (USA).

\sloppy
\raggedright
\clearpage
\printbibliography[title={References},heading=bibintoc]

\end{document}